\documentclass[prl,twocolumn,showpacs,superscriptaddress]{revtex4}
\usepackage{graphicx}
\usepackage{amsfonts}
\usepackage{amsmath}
\usepackage{amssymb}
\usepackage{bm}

\usepackage{amsmath}
\usepackage{color}
\usepackage{textcomp}
\usepackage[colorlinks=true, letterpaper=true, pdfstartview=FitV, linkcolor=blue, citecolor=blue, urlcolor=blue]{hyperref}
\begin{document}

\title{Topological phase transition between distinct Weyl semimetal states in MoTe$_2$}

\author{Anmin Zhang}
\thanks{These authors contributed equally to this work.}
\affiliation{School of Physical Science and Technology, Lanzhou University, Lanzhou 730000, 
China}
\affiliation{Department of Physics, Renmin University of China, Beijing 100872, China}
\author{Xiaoli Ma}
\thanks{These authors contributed equally to this work.}
\author{Changle Liu}
\author{Rui Lou}
\author{Yimeng Wang}
\author{Qiaohe Yu}
\author{Yiyan Wang}
\author{Tian-long Xia}
\author{ Shancai Wang}
\affiliation{Department of Physics, Renmin University of China, Beijing 100872, China}
\author{Lei Zhang}
\affiliation{Anhui Key Laboratory of Condensed Matter Physics at Extreme Conditions, High Magnetic Field Laboratory, Chinese Academy of Sciences, Hefei 230031, China}
\author{Xiaoqun Wang}
\affiliation{Department of Physics and Astronomy, Shanghai Jiao Tong University, Shanghai 200240, China}
\author{Changfeng Chen}
\affiliation{Department of Physics and Astronomy, University of Nevada, Las Vegas, Nevada 89154, USA}
\author{Qingming Zhang}
\email{qmzhang@ruc.edu.cn}
\affiliation{Department of Physics, Renmin University of China, Beijing 100872, China}
\affiliation{Beijing National Laboratory for Condensed Matter Physics, Institute of Physics, Chinese Academy of Sciences, Beijing 100190, China}
\affiliation{School of Physical Science and Technology, Lanzhou University, Lanzhou 730000, China}

\date{\today}

\begin{abstract}
 We present the first experimental evidence of an intriguing phase transition between 
 distinct topological states in the novel type-II Weyl semimetal MoTe$_2$. We observe 
 anomalies in Raman phonon frequencies and linewidths as well as electronic quasielastic 
 peaks around 70 K, which together with structural, thermodynamic measurements and  
 electron-phonon coupling (EPC) calculation demonstrate a temperature-induced transition 
 between two topological phases previously identified by contrasting spectroscopic 
 measurements. Analysis of experimental data suggests electron-phonon coupling as the 
 main driving mechanism for the change of key topological characters in the electronic 
 structure of MoTe$_2$. We also find the phase transition sensitive to sample conditions 
 distinguished by synthesis methods. These discoveries of temperature and material 
 condition dependent topological phase evolution and transition in MoTe$_2$ advance 
 fundamental understanding of the underlying physics and enable an effective approach to 
 tuning Weyl semimetal states for technological applications.

\end{abstract}


\maketitle
Recent years have seen the dramatic rise of a new class of quantum materials whose electronic states exhibit symmetry protected topological orders \cite{1,2}. Such electronic states are insensitive to local decoherence processes, thus offering great promise for constructing quantum computing and high-speed electronic and spintronics devices. Distinct topological states, such as topological insulators, Dirac semimetals, and type-I and type-II Weyl semimetals, have been theoretically \cite{1,2,3,4,5,6,7,8} proposed and experimentally \cite{9,10,11,12,13,14,15,16,17,18} realized in real materials. A central task in this research field is to unravel the material and environment (e.g., pressure, temperature, etc.) conditions conducive to the existence of topologically ordered phases. To this end, it is essential to be able to induce and control the phase transition that allows an effective manipulation of the unique properties of the topological states. Significant progress has been made in understanding transitions between topologically trivial and nontrivial states. Recent theoretical studies have shown that strain \cite{19,20}, phonon \cite{21,22} and/or disorder \cite{23} can induce topological phase transitions that greatly influence electronic states and properties. For instance, when a topological transition occurs, topological surface states dramatically change \cite{8,9,10,11,12,13}, greatly impacting the novel transport behaviors \cite{15,16,17,18,24}, and phonon modes strongly coupled to electrons may also behave anomalously \cite{25,26}. A recent experiment revealed that a structural transition can act as a switch of the topological phase transition \cite{27}. Meanwhile, however, transitions between distinct topologically ordered states have remained largely unexplored, especially on the experimental front.

Molybdenum ditelluride (MoTe$_2$), a type-II Weyl semimetal (WSM) \cite{19,20,28}, offers an excellent platform to probe distinct topological phases and possible transitions among them. MoTe$_2$ has three structural phases: \emph{2H} (hexagonal, space group P63/mmc), \emph{1T'} (monoclinic, P21/m) and $T_d$ (orthorhombic, Pnm21) \cite{29}. Topological surface states have been observed in $T_d$-MoTe$_2$ at low temperatures \cite{30,31,32,33,34,35}. Angle-resolved photoemission (ARPES) measurements \cite{33} suggested that MoTe$_2$ specimen grown by a flux method harbors an electronic structure containing four Weyl points (WPs), and this conclusion was supported by electronic band calculations \cite{20}. On the other hand, contrasting ARPES experiments \cite{34,35} observed eight WPs in MoTe$_2$ grown by a chemical vapor transport method, and the results are also supported by calculations \cite{19,35}. The diverging theoretical results likely stem from using the different lattice constants measured at different temperatures \cite{19,20}, while the variation of the experimental results suggests that the nature of the topological states are highly sensitive to sample conditions. These results imply that a topological phase transition may occur in MoTe$_2$ \cite{19,20}, but experimental evidence is still lacking. The objective of our present work is to seek and establish clear experimental signatures of the transition between distinct Weyl semimetal phases, thus unifying the seemingly inconsistent ARPES results and resolving the nature of the topological states in MoTe$_2$, which is crucial to understanding fundamental physics of these novel topological quantum materials and their exotic properties, such as topological superconductivity \cite{36,37,38,39,40,41,42,43}.

 In this work, we examine the topological states in MoTe$_2$ by performing Raman 
 scattering, structural and thermodynamic measurements on crystals grown using different 
 synthesis methods (The crystal-growth and measurement methods can be found in 
 Supplemental Material \cite{SM}). Our results reveal that several phonon modes show 
 strong EPC effects. The frequencies, widths and q factors of these phonon modes exhibit 
 clear anomalies at $\sim$ 70 K in MoTe$_2$ crystals grown by a flux method (hereafter 
 referred to as Flux MoTe$_2$), but, surprisingly, not in the crystals grown by a chemical 
 vapor transport method (CVT MoTe$_2$). The intensities of the low frequency quasielastic 
 peak (QEP), originating from the electronic Raman response, also show a minimum at 70 K 
 in flux MoTe$_2$ but not in CVT MoTe$_2$. These anomalies are further corroborated by our 
 transport measurements. All these observations together with EPC calculation establish a 
 clear case of sample dependent topological characters of the electronic structure and an 
 intriguing phase transition in flux MoTe$_2$ between distinct Weyl semimetal states. Raman 
 scattering results together with x-ray diffraction (XRD), transport, specific heat data and EPC 
 	calculation
 consistently demonstrate that the transition is an electronic/topological phase transition 
 driven by the strong EPC effect. These results lead to a phase diagram that helps distinguish 
 different topological ground states of MoTe$_2$ under different material and temperature 
 conditions. 
 

\begin{figure}[tb]
\includegraphics[angle=0,width=8cm]{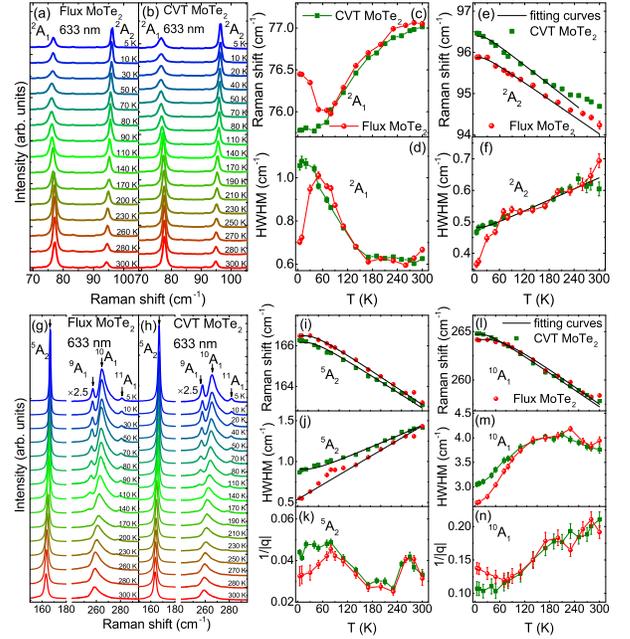}
\caption{(Color online) Anomalies in phonon Raman spectra. (a), (b), (g) and (h) 
Temperature evolution of the  $^2A_1, ^2A_2, ^5A_2$  and $^{10}A_1$ modes in Flux 
MoTe$_2$ and CVT MoTe$_2$. (c)-(f), (i)-(n) Temperature dependence of peak positions, 
linewidths and asymmetry factor \emph{q} of these phonon modes.} 
\label{fig1}
\end{figure}

\begin{figure}[tb]
\includegraphics[angle=0,width=8cm]{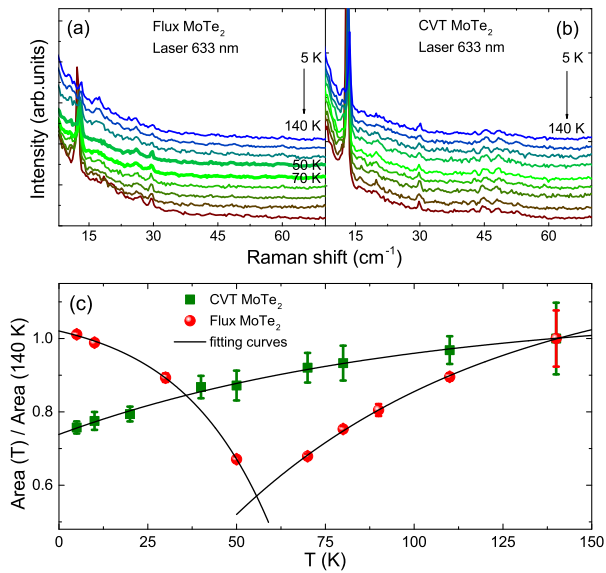}
\caption{(Color online) Anomalies in quasielastic peaks at low wavenumbers. (a), (b) 
Low-frequency quasielastic peaks at various temperatures in Flux MoTe$_2$ and CVT 
MoTe$_2$. (c) Temperature dependence of the normalized intensities of quasielastic peaks. 
Black lines in (c) are fitted using the exponential functions ($I=I_0+I_1e^{\frac{-T}{\tau}}$ 
where $I_0, I_1$, $\tau$ are parameters.). The phonons below 15 cm-1 are shear 
	mode\cite{ramanNL} with $A_{1}$ symmetry. So the intensities  are strongly 
	affected 
	by the angle between the polarization of the incident light and the crystallographic axis. The 
	shear mode 
	shows different intensities in (a) and (b) for the different angles. }
\label{fig2}
\end{figure}

Low-temperature $T_d$-MoTe$_2$ hosts 33 Raman modes, among which 17 modes visible on 
the \emph{ab} plane have been well assigned \cite{44, ramanNL, ramanNC, ramanAPL}. 
Here we focus on the modes with stronger intensities. These modes can be divided into two 
groups: those with symmetric peak profiles (Fig. 1(a) and 1(b)), and others with asymmetric 
profiles (Fig. 1(g) and 1(h)). The temperature evolution of the $^2A_1$ and $^2A_2$ modes, 
which are the vibrations dominated by Te atoms along the b and a axes respectively 
\cite{44}, is shown in Fig. 1(a) and 1(b). In Fig. 1 (c)-(f), we show the temperature 
dependence of their frequencies and linewidths extracted by a Lorentzian fitting (an example 
of fitting is shown in Supplemental Material, Fig. S1 \cite{SM}).

The $^2A_1$ mode exhibits an unusual softening in frequency and broadening in width with 
decreasing temperature (Fig. 1(c) and 1(d)). This is in stark contrast to the normal hardening 
and narrowing of the $^2A_2$ mode related to the anharmonic phonon coupling \cite{45} 
(Fig. 1(e) and 1(f)), and can be attributed to the EPC (see the Supplemental Material, Sec. III 
	and Sec. X \cite{SM}). Consistently, the anomalous broadening in linewidth can be also well 
understood within the EPC picture (Fig. 1(d)). In the EPC theory \cite{46}, the real part of the 
self-energy represents the frequency shift and the imaginary part gives the phonon 
broadening due to its interaction with the electronic continuum. Both parts are driven 
toward constant values by the Fermi-Dirac distribution function approaching zero 
temperature, as seen in CVT MoTe$_2$ (Fig. 1(c) and 1(d)). Surprisingly, the flux MoTe$_2$ 
follows a softening tendency at higher temperatures but upon cooling exhibits a dramatic 
upturn at $T_t \sim$ 70 K (Fig. 1(c) and 1(d)). This anomalous behavior clearly signals 
changes in the underlying phonon or electronic structures.

The temperature evolution of the asymmetric $^5A_2$ and $^{10}A_1$ modes is presented in Fig. 1(g) and 1(h). In the phonon softening picture, this asymmetry reflects the EPC  described by the Fano formula \cite{46}. The parameters by Fano fitting and their temperature dependence are summarized in Fig. 1(i)-(n) (The fitting details are shown in Supplemental Material, Fig. S1 \cite{SM}).

For the $^5A_2$ and $^{10}A_1$ modes, the anomalies in linewidth and 1/$|q|$ at $\sim$ 70 K can be clearly seen in flux MoTe$_2$ but not in CVT MoTe$_2$. The anomalies unambiguously point to a phase transition in flux MoTe$_2$. In other words, flux MoTe$_2$ and CVT MoTe$_2$ share the same phase at high temperatures, but fall into different phases below 70 K. The observation that there is no anomaly in the lattice constant and/or specific heat at the transition temperature (Fig. 3(c)) suggests that this phase transition is not driven by structural changes (see below for further discussions). We have repeated Raman scattering measurements on different batches of crystals, and the results show perfect repeatability (see Supplemental Material, Fig. S3 \cite{SM}).

The phase transition at 70 K is further evidenced by electronic Raman response (ERS), which is an inelastic light scattering process by band electrons and includes the influence of low-energy single-particle excitations and high-energy collective plasmon excitations. The transferred energies in this process would be very small if the involved photons and band electrons have small momenta and small Fermi velocities, respectively. In such a case, electronic Raman scattering manifests itself as a quasielastic peak (QEP) \cite{47}. The low-energy QEPs can be clearly seen in both the flux and CVT MoTe$_2$ samples (Fig.2(a) and 2(b)). However, the temperature dependences of QEPs are very different for the two samples. In CVT MoTe$_2$, the QEP intensity monotonously declines with decreasing temperature and smoothly passes through the transition temperature; meanwhile, the QEP in flux MoTe$_2$ exhibits a clear minimum at $T_t\approx$ 70 K. This difference suggests two distinct phases at low temperatures for the two samples, in agreement with the findings from the phonon spectra. And the anomaly in QEP points to a phase transition of the electronic states since QEP is contributed purely by electrons. This assessment is further supported by our transport measurements. We extracted electronic concentration $n_e$ and the mobilities of both electrons and holes from the two-carrier model analysis of magneto-resistivity $\rho_{xx}$ and Hall resistivity $\rho_{xy}$. The results show a sharp upturn below 70 K in Flux MoTe$_2$ but smoothly evolve with temperature in CVT MoTe$_2$ (Supplemental Material, Fig. S4 \cite{SM}). This observation confirms the phase transition of the electronic states in Flux MoTe$_2$.

\begin{figure}[tb]
	\includegraphics[width=8cm]{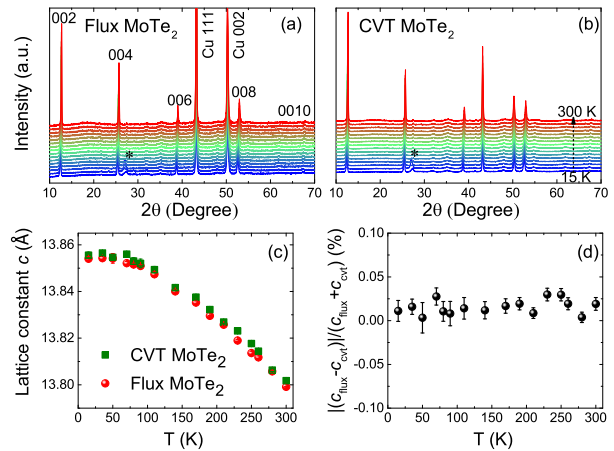}
	\caption{(Color online) Nearly identical structural parameters in two samples. (a), (b) XRD patterns at various temperatures in Flux MoTe$_2$ and CVT MoTe$_2$. (c) Temperature dependence of lattice constant \emph{c}. (d) Relative differences of \emph{c} between two samples at various temperatures, which remain less than 0.05$\%$ at all temperatures. The broad peaks at low temperatures marked by * in (a) and (b) are instrumental signals (see Supplemental Material, Fig. S6 \cite{SM}).}
	\label{fig3}
\end{figure}

To assess the origin of the  phase transition, we further performed XRD measurements on the two samples. The XRD patterns at various temperatures are shown in Fig. 3(a) and 3(b). The preferred-orientation effect of layered materials makes the (00L) peaks much stronger than others. The extracted lattice constants (Fig. 3(c)) indicate that MoTe$_2$ shows a negative thermal expansion along the \emph{c} axis, in agreement with previous studies \cite{35}. The lattice constant \emph{c} has a smooth temperature evolution and tends to be saturated below 100 K. This result supports the conclusion that the softening of the $^2A_1$ mode and the anomalies around $T_t$ are not related to any structural change. The relative difference of \emph{c} between the two samples is negligibly small at less than 0.05$\%$ at all temperatures (Fig. 3(d)). Thus, we can conclude that the phase transition at 70 K in flux MoTe$_2$ is an electronic phase transition rather than a structural one, which is corroborated by the absence of any anomaly around $T_t$ in specific heat of flux MoTe$_2$ (Supplemental Material, Fig. S2 \cite{SM}). All these results point to the conclusion that an electronic topological phase transition occurs in flux MoTe$_2$ but not in CVT MoTe$_2$.

To further characterize the material dependence of the topological phase transition in MoTe$_2$, we have measured the resistivity as a function of temperature from 2.5 K to 300 K on two batches of MoTe$_2$ crystals, two synthesized using the flux method and two using the CVT method (Fig. 4(a) and Supplemental Material, Fig. S5 \cite{SM}). Remarkably, the residual resistivity ratio (RRR), defined as the ratio of resistivity values at room temperature (300 K) and in the low-temperature limit , of flux MoTe$_2$ is an order of magnitude larger than that of CVT MoTe$_2$. Such a large disparity in RRR, also reported in previous studies \cite{30,31,36,37,38,39,48,49}, implies some degree of disorder or inhomogeneity in CVT MoTe$_2$. From our sample- and temperature-dependent data, we have constructed a phase diagram in the T-RRR phase space for the two topological phases (Fig. 4(b)), containing eight and four WPs, respectively, which have been observed in separate ARPES measurements in the MoTe$_2$ crystals grown by CVT and flux methods \cite{30,31,32,33,34,35}. Our Raman data are consistent with the ARPES observations as illustrated in the T-RRR phase diagram, where the phase at higher temperatures is in one topological state (TP I, 8-WP), while the phase in the high-RRR and low-temperature region is in another topological state (TP II, 4-WP). Our results show that upon cooling the high-RRR flux MoTe$_2$ sample undergoes a transition from the TP I to TP II Weyl semimetal state; meanwhile, the low-RRR CVT MoTe$_2$ sample remains in the TP I state throughout the entire temperature range. These results explain the previously reported divergent topological characters observed in different MoTe$_2$ crystals \cite{19,20,33,34,35,36}. The sample-sensitive temperature-induced transition highlights the material (RRR) and environment (temperature) dependence of the topological state in MoTe$_2$. Further studies are required to map out the full phase boundary between the TP I and TP II states; one may also establish phase diagrams in other material and environment parameter spaces.

\begin{figure}[tb]
\includegraphics[width=8cm]{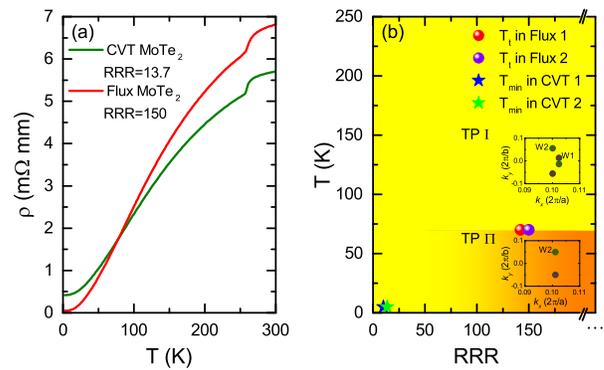}
\caption{(Color online) R-T curves and T-RRR phase diagram. (a) R-T curves of a Flux MoTe$_2$ and a CVT MoTe$_2$. (b) Temperature-RRR phase diagram of MoTe$_2$. In the low-RRR region, the star symbols denote the data points taken on two CVT MoTe$_2$ samples at the lowest temperature ($T_{min}$=5 K) in our measurements, where no signs of a temperature-induced phase transition have been detected. In the high-RRR region, the sphere symbols denote the transition temperatures ($T_t$) given by the phonon Raman spectra obtained on two flux MoTe$_2$ samples. The Insets in (b) are schematic band structures in the two topological phases \cite{19,20}.} 
\label{fig4}
\end{figure}

The key topological character of a Weyl semimetal is the nodal structure of its electronic 
bands, where the two-fold degenerate bands cross with a linear dispersion. The number and 
distribution of such Weyl points (WPs) define the topological properties crucial to 
fundamental understanding and practical applications of these novel materials. The WPs in 
MoTe$_2$ are generally divided into two groups, namely W1 WPs that are closely positioned 
in the momentum space and W2 WPs that are well separated. It was shown theoretically that 
a strain of 0.1$\%$ along the \emph{a} axis of MoTe$_2$ can tune the appearance of W1 
WPs \cite{20}. According to the temperature dependence of the lattice constant \emph{a}, a 
temperature change over 70 K (from 0 K to $T_t$) would induce a strain over 0.1$\%$ in 
MoTe$_2$ \cite{35}, making a topological phase transition induced by the thermal expansion 
plausible. Our experiments show, however, that the lattice constants of flux MoTe$_2$ and 
CVT MoTe$_2$ remain very close (less than 0.05$\%$ difference) in the entire studied 
temperature range, yet the transition only occurs in the flux samples. This result rules out 
the change in lattice constant as the origin of the observed topological phase transition. 
Meanwhile, temperature influence on $E_F$ may trigger a Lifshitz transition. However, it has 
been shown that a temperature change from 0 K to $T_t$ (70 K) is too small to bring about an 
observable modification in $E_F$ sufficient to drive a topological transition in MoTe$_2$ 
\cite{48}, while the change in the ratio between the hole and electron densities ($n_h/n_e$) 
near $T_t$ \cite{39,48} may reflect the reconstruction of Fermi surface induced by the 
topological transition. Finally, a strong EPC in MoTe$_2$ has been revealed in our 
experiments. The EPC can simultaneously alter the phononic energy and lifetime and the 
electronic structure. A topological transition induced by EPC has been theoretically studied 
\cite{21,22}. In our paper, we also made the EPC
	calculation (see details in Supplemental
	Material \cite{SM}, Sec. X), which shows that the EPC contribution to the electronic 
	self-energy generally increases with temperature because of the fast rising in the phonon 
	occupation number, and the EPC induced changes in the electronic structure could close the 
	small gap between two bands, which eventually cross and form W1 WPs, causing a 
	topological phase transition from one phase (TP II) hosting four WPs to another phase (TP I) 
	containing eight WPs. When the topological phase transition occurs, the phonon 
	linewidth reverses as well. This  behavior can serve as a good indicator of electronic band 
	inversions \cite{26}, thus explaining the anomaly of the linewidths of the $^2A_1$ mode 
	observed in MoTe$_2$.

Our results indicate a transition of the topological ground state from the TP I phase with eight WPs at temperatures above 70 K to the TP II phase with four WPs at lower temperatures in the high-RRR flux MoTe$_2$ crystals. Meanwhile, the low-RRR CVT MoTe$_2$ remains in the TP I phase in the entire temperature region from 300 K down to 5 K, which is lower than in recent ARPES experiments (10 and 6-20 K) \cite{34,35}. This suggests that disorder may have caused substantial changes in the topological ground state of CVT MoTe2 by modifying the electronic self-energies \cite{23}. It indicates that the EPC in CVT MoTe$_2$ is significantly overcome by the disorder effect and can no longer induce a topological phase transition (TP II to TP I) as observed in flux MoTe$_2$. These contrasting results raise important questions about fundamental interactions in MoTe$_2$.

We have shown Raman, structural and transport measurements that demonstrate a  
	temperature-induced electronic transition in flux MoTe$_2$ and the EPC calculation 
	indicates this  transition is a topological phase transition between  distinct Weyl 
	semimetal states. These results, combined with our and existing ARPES data, identify a 
topological 
ground-state transition from a TP I phase hosting eight WPs at temperatures above 70 K to a 
TP II phase containing four WPs at lower temperatures. Our study also shows that MoTe$_2$ 
crystals obtained by a chemical vapor transport method remains in the TP I phase down to 5 
K without any sign of a phase transition. Based on these findings, we have constructed a 
temperature-RRR phase diagram that reconciles the divergent views on the topological 
characters of the electronic structure of MoTe$_2$. The present results have broad 
implications for major topological properties such as the surface Fermi arc structures and 
transport behaviors; they also raise fundamental questions on the roles of key physical 
processes and material conditions in determining the properties of topological materials. Our 
experiments and calculation suggest strong EPC in MoTe$_2$ as the driving mechanism for 
the observed 
topological phase transition, highlighting a major influence of EPC effects in topological 
materials. Our results also indicate that disorder may have a significant role in impeding a 
topological phase transition in CVT MoTe$_2$. These insights advance and enrich 
fundamental understanding of Weyl semimetals and pave the way for further research to 
unveil new physics in this class of novel materials.

\section*{Acknowledgements} We thank Shanshan Sun for assisting with specific heat measurements. This work was supported by the Ministry of Science and Technology of China (2016YFA0300504 and 2017YFA0302904) and the NSF of China (11774419, 11474357, 11604383 and 11574391). A.M.Z and T.L.X were supported by the Fundamental Research Funds for the Central Universities and the Research Funds of Renmin University of China (18XNLG23 and 14XNLQ07).

\end{document}